\documentclass[superscriptaddress,showpacs,showkeys,preprintnumbers,aps,nofootinbib,prd]{revtex4-1}
\usepackage{graphicx}
\usepackage{epsfig}
\usepackage{amsmath}
\usepackage{amsfonts}
\usepackage{amssymb}

\begin{document}
\title{\textbf{Inflation with a quartic potential in the framework of Einstein-Gauss-Bonnet gravity}}

\author{Ekaterina~O.~Pozdeeva}
\email{pozdeeva@www-hep.sinp.msu.ru}
\affiliation{Skobeltsyn Institute of Nuclear Physics, Lomonosov Moscow State University,
 Leninskie Gory 1, Moscow  119991,  Russia}

\author{Mayukh Raj Gangopadhyay}
\email{mayukh@ctp-jamia.res.in}
 \affiliation{Centre
for Theoretical Physics, Jamia Millia Islamia, New Delhi 110025,
India}

\author{Mohammad~Sami}
\email{msami@jmi.ac.in}
\affiliation{International Center for Cosmology, Charusat University, Anand 388421, Gujarat, India}
\affiliation{Maulana Azad National Urdu
University, Gachibowli, Hyderabad 500032, India}
\affiliation{Institute for Advanced Physics and Mathematics, Zhejiang University of Technology,\\ Hangzhou 310032, China}
\affiliation{Center for Theoretical Physics, Eurasian National University, Astana 010008, Kazakhstan}

\author{Alexey~V.~Toporensky}
\email{atopor@rambler.ru}
\affiliation{Sternberg Astronomical Institute, Lomonosov Moscow State University, Universitetsky pr.~13, Moscow 119991, Russia}
\affiliation{Kazan Federal University, Kremlevskaya Street~18, Kazan 420008, Russia}

\author{Sergey~Yu.~Vernov}\email{svernov@theory.sinp.msu.ru}
\affiliation{Skobeltsyn Institute of Nuclear Physics, Lomonosov Moscow State University,
 Leninskie Gory 1, Moscow  119991,  Russia}

\begin{abstract}
 We investigate inflationary dynamics in the framework of the
Einstein-Gauss-Bonnet gravity. In the model under consideration, the inflaton
field is nonminimally coupled to the Gauss-Bonnet curvature invariant, so that
the latter appears to be dynamically important. We consider a quartic potential
for the inflaton field, in particular, the one asymptotically connected to the
Higgs inflation, and a wider class of coupling functions not considered in the
earlier work. Keeping in mind the observational bounds on the parameters --- the
amplitude of scalar perturbations $A_s$, spectral index $n_s$ and
tensor-to-scalar ratio $r$, we demonstrate that the model a quartic potential
and the proposed coupling function is in agreement with observations.
\end{abstract}




\maketitle

\section{Introduction}

The current observations on Cosmic Microwave Background put tight constraints on models of inflation~\cite{Planck2018}. The popular models with quadratic and quartic potentials~\cite{Linde:1983gd}, that played an important role in the history of inflationary model building, already turn out to be
incompatible with observations, as the tensor-to-scalar ratio of perturbations which these models predict is too large~\cite{Mukhanov:2013tua,Kallosh:2013pby,Creminelli:2014fca}.
If one adheres to  General Relativity, the field potential  should be rather gently sloping in order to match with observations~\cite{Mukhanov:2013tua,Ventury2015,Maity}. These restrictions however, can drastically be softened in modified theories of gravity~\cite{Kallosh:2013pby,predictions,Barvinsky:1994hx,Cervantes-Cota1995,BezrukovShaposhnikov,EOPV2014,Koshelev:2020xby}.

A well-known example is provided by the Higgs field as an inflaton~\cite{BezrukovShaposhnikov}, which due to its large self-coupling gives rise
to the large amplitude of scalar perturbations in the framework of the standard theory of gravity\footnote{The tensor-to-scalar ratio of perturbations is also large in this case which also applies to the model-based upon quadratic potential. }.
However, a nonminimal coupling of the Higgs field with the curvature can significantly reduce the amplitude of scalar perturbations bringing the model within the observational bounds. Unfortunately, the price for this is paid by the large numerical value of the dimensionless constant of the nonminimal coupling which sounds unnatural\footnote{The numerical value of nonminimal coupling constant turns out to be about 50000 which is rather large compared to  unity for a dimensionless fundamental constant.}. This motivates a search for more complicated
scenarios of inflation, in particular, the one obtained by adding the term proportional to $R^2$
to the Einstein-Hilbert Lagrangian. Another possibility is provided by the $\alpha$-attractor formalism which also allows lowering of the tensor-to-scalar ratio of perturbations down to the ones consistent with observation~\cite{Kallosh:2013hoa,Elizalde:2015nya}.

We hereby consider a scenario that uses a particular form of quadratic gravity, namely,
 the Gauss-Bonnet (GB) term added to the Einstein-Hilbert action.  Being a total derivative, this term alone does not contribute to equations of motion, it, however,  becomes dynamically important if coupled with a function of the scalar field, $\xi(\phi)$. This might have
important implications for inflation as well as for late time acceleration. On the other hand, the GB term arises naturally in the string  theory framework as a quantum correction to the Einstein-Hilbert action~\cite{Antoniadis:1993jc,Torii:1996yi,Kawai1998,Hwang:2005hb,Tsujikawa:2006ph,ss,Sami:2005zc,Cognola:2006sp}.

A plethora of inflationary models with the GB term~\cite{Soda2008,Guo:2009uk,Guo,Jiang:2013gza,Koh,vandeBruck:2015gjd,JoseMathew,vandeBruck:2016xvt,Koh:2016abf,Nozari:2017rta,Armaleo:2017lgr,Chakraborty:2018scm,Yi:2018gse,Odintsov:2018,Fomin:2019yls,Kleidis:2019ywv,Rashidi:2020wwg,Odintsov:2020sqy,Pozdeeva:2020shl}, including the ones with cosmological attractor  constructs~\cite{Nozari:2017rta,Pozdeeva:2020shl} have been discussed in the literature.
The most actively studied models with GB coupling involves the function $\xi$ inversely proportional to
the scalar field potential~\cite{Guo:2009uk,Guo,Jiang:2013gza,Koh:2016abf,Yi:2018gse,Odintsov:2018,Kleidis:2019ywv,Rashidi:2020wwg,Pozdeeva:2020shl}.
Considering this specific form of the coupling function, it has been demonstrated that models based upon quadratic and quartic potentials could be rescued: the GB term reduces the tensor-to-scalar ratio $r$ to appropriate values consistent with observation~\cite{Guo,Yi:2018gse}. Interestingly, the GB term does not change the value of  the spectral index $n_s$ for the mentioned choice of the coupling function.
To the best of our knowledge, there is no physical reason for such a choice of the function $\xi(\phi)$.
The particular form of coupling function $\xi(\phi)$ is motivated by the consideration of simplicity. It would be plausible to enlarge the class of  coupling functions and check the observational viability of the respective models.

In the present paper, we investigate a wider class of theories involving  GB coupling with coupling function $\xi(\phi)$ not related to the inflaton potential,
and show that the above-mentioned tensions with observation can
be removed. In this framework, it is possible to reduce the scalar amplitude such that the Higgs field coupled
to GB term could provide reasonable perturbation parameters compatible with current observational limits. Note that we consider the minimal coupling between the scalar field and the scalar curvature.

A useful tool that allows us to obtain different inflationary models with the GB term is provided by the construction of an effective potential. This function has been proposed to study stability of de Sitter solutions in models with nonminimal coupling~\cite{Skugoreva:2014gka, Pozdeeva:2016cja} and generalized to the models with the GB term~\cite{Pozdeeva:2019agu}. In this paper, we use the formalism of the effective potential for investigations of inflationary
parameters in the models under consideration.

\section{The slow-roll parameters and the leading order equations}

In what follows, we shall consider the modified gravity model with the GB term,~\cite{Guo}:
\begin{equation}
\label{action1}
S=\int d^4x\sqrt{-g}\left[UR-\frac{g^{\mu\nu}}{2}\partial_\mu\phi\partial_\nu\phi-V(\phi)-\frac{\xi(\phi)}{2}\mathcal{G}\right],
\end{equation}
where $U$ is a positive constant, the functions $V(\phi)$, and $\xi(\phi)$ are differentiable ones, and $\mathcal{G}$ is the Gauss-Bonnet term: $\mathcal{G}=R_{\mu\nu\rho\sigma}R^{\mu\nu\rho\sigma}-4R_{\mu\nu}R^{\mu\nu}+R^2$.

Application of the variation principle leads to the following system of equations in the spatially flat Friedmann universe~\cite{vandeBruck:2015gjd,Pozdeeva:2019agu}:
\begin{eqnarray}
               12UH^2 &=& \dot\phi^2+2V+24\dot{\xi}H^3,  \label{Equ0} \\
                4U\dot{H} &=& {}-\dot\phi^2+4\ddot{\xi}H^2+4\dot{\xi}H\left(2\dot{H}-H^2\right), \label{EquH}\\
                \ddot{\phi}&=&{}-3H\dot{\phi}-V'-12\xi'H^2\left(\dot{H}+H^2\right), \label{Equphi}
\end{eqnarray}
where  $H=\dot{a}/a$ is the Hubble parameter, $a(t)$ is the scale factor, dots and primes denote the derivatives with respect to the cosmic time  $t$ and  the scalar field $\phi$, respectively.

During inflation $H(t)$ is always finite and positive, therefore, it is possible to use the dimensionless parameter
$N=\ln(a/a_{e})$,
where $a_{e}$ is a constant, as a new measure of time\footnote{Note that in many papers~\cite{Guo,vandeBruck:2015gjd,Odintsov:2020sqy,Geng:2017mic} $N=0$ corresponds to the beginning of inflation, whereas we fix $N=0$ at the end of inflation.  Also, there is an alternative definition of the e-folding number: $\tilde N=-\ln(a/a_e)$, see~\cite{Mukhanov:2013tua,Koh:2016abf,Pozdeeva:2020shl}. }.

Following Refs.~\cite{Guo,vandeBruck:2015gjd}, we consider the slow-roll parameters:
\begin{eqnarray*}
  \epsilon_1 &=&{}-\frac{\dot{H}}{H^2}={}-\frac{d\ln(H)}{dN},\qquad \epsilon_{i+1}= \frac{d\ln|\epsilon_i|}{dN},\quad i\geqslant 1, \\
  \delta_1&=& \frac{2}{U}H\dot{\xi}=\frac{2}{U}H^2\xi'\frac{d{\phi}}{dN},\qquad \delta_{i+1}=\frac{d\ln|\delta_i|}{dN},\quad i\geqslant 1,
\end{eqnarray*}
where we use  ${d}/{dt}=H\, {d}/{dN}$.  The slow-roll approximation requires $|\epsilon_i|\ll1$ and $|\delta_i|\ll1$. We fix $a_e$ by the condition $\epsilon_1=1$.

The slow-roll conditions $\epsilon_1\ll 1$, $\epsilon_2\ll 1$, $\delta_1\ll 1$, and $\delta_2\ll 1$ allow to simplify Eqs.~(\ref{Equ0})--(\ref{Equphi}).
Indeed,
\begin{equation}
\label{delta2equ}
\delta_2=\frac{\dot\delta_1}{H\delta_1}=\frac{2\ddot\xi}{U\delta_1}-\epsilon_1,
\end{equation}
so from $|\delta_2|\ll 1$ and $|\epsilon_1|\ll 1$ it follows $|\ddot\xi|\ll |H\dot\xi|$.
Using $|\delta_1|\ll 1$ and $|\delta_2|\ll 1$, we obtain from  Eqs.~(\ref{Equ0}) and (\ref{EquH}):
\begin{eqnarray}
               &&12UH^2 \simeq \dot\phi^2+2V,  \label{Equ0slr1} \\
                &&4U\dot{H} \simeq{}-\dot\phi^2-4\dot{\xi}H^3={}-\dot{\phi}\left(\dot{\phi}+4\xi'H^3\right). \label{EquHslr1}
\end{eqnarray}
Using,
\begin{equation*}
\epsilon_1={}-\frac{\dot H}{H^2}\simeq \frac{\dot\phi^2}{3(\dot\phi^2+2V)}+\frac{1}{2}\delta_1\ll 1,
\end{equation*}
we obtain $\dot\phi^2 \ll 2V$, and Eq.~(\ref{Equ0slr1}) takes the following form
\begin{equation}
\label{Equ0slr2}
6UH^2 \simeq V.
\end{equation}
Taking the time derivative of this equation and using Eq.~(\ref{EquHslr1}), we get
\begin{equation}
\label{Equphislr1}
    \dot{\phi}\simeq{}-\frac{V'}{3H}-4\xi'H^3.
\end{equation}
Substituting this relation to Eq.~(\ref{Equphi}), we get that $|\ddot{\phi}|\simeq|12\xi'H^2\dot{H}|\ll|12\xi'H^4| $.
Thus, the slow-roll conditions result to
 \begin{equation*}
 \dot\phi^2\ll V, \quad |\ddot{\phi}|\ll |12\xi'H^4|,\quad 2|\dot{\xi}|H\ll U,\quad |\ddot{\xi}|\ll|\dot{\xi}|H\,,
 \end{equation*}
 so, the leading order equations in the slow-roll approximation have the following form:
 \begin{eqnarray}
               H^2&\simeq&\frac{V}{6U}\,, \label{Equ0lo}\\
               \dot{H}&\simeq&{}-\frac{\dot\phi^2}{4U}-\frac{\dot{\xi}H^3}{U}\,, \label{EquHlo}\\
                \dot{\phi}&\simeq&{}-\frac{V'+12\xi'H^4}{3H}. \label{Equphilo}
\end{eqnarray}

In the following section, we briefly discuss the effective potential formalism to be used for analysing the inflationary dynamics.

\section{The effective potential and inflationary scenarios}

\subsection{The slow-roll approximation}
\label{slow-roll approximation}
To analyze the stability of de Sitter solutions in model (\ref{action1}) the effective potential has been proposed in Ref.~\cite{Pozdeeva:2019agu}:
\begin{equation}
\label{Veff}
V_{eff}(\phi)={}-\frac{U^2}{V(\phi)}+\frac{1}{3}\xi(\phi).
\end{equation}

The effective potential is not defined in the case of $V(\phi)\equiv 0$, but  inflationary scenarios are always unstable in this case~\cite{Hikmawan:2015rze} (see also~\cite{Chakraborty:2018scm}).
In this paper,
 we consider inflationary scenarios with positive potentials only: $V(\phi)> 0$ during inflation.
The effective potential characterizes existence and stability of de Sitter solutions completely.
It is however not enough to fully characterize quasi-de Sitter inflationary stage, and the potential
$V(\phi)$ enters into expressions of the inflationary parameters (see below) as well. Nevertheless,
keeping the effective potential in the corresponding formulae
would be helpful, as we will see soon.

Using Eqs.~(\ref{EquHlo}) and (\ref{Equphilo}), we get that the functions $H(N)$ and $\phi(N)$ satisfy the following leading order equations:
\begin{eqnarray}
               \frac{d{H}}{dN}&\simeq&{}-\frac{H}{U}V'V_{eff}'\,, \label{EquHloN}\\
               \frac{d{\phi}}{dN}&\simeq&{}-2\frac{V}{U}V_{eff}'. \label{EquphiloN}
\end{eqnarray}

In terms of the effective potential the slow-roll parameters are as follows:
\begin{equation}
\label{slrVeffe}
    \epsilon_1={}-\frac{1}{2}\frac{d\ln(V)}{dN}=\frac{V'}{U}V_{eff}'\,,
    \end{equation}
\begin{equation}
    \epsilon_2={}-\frac{2V}{U}V_{eff}'\left[\frac{V''}{V'}+\frac{V_{eff}''}{V_{eff}'}\right]
    ={}-\frac{2V}{U}V_{eff}'\left[\ln(V'V_{eff}')\right]'\,,
\end{equation}
\begin{equation}
  \delta_1= {}-\frac{2V^2}{3U^3}\xi'V_{eff}'\,,
  \end{equation}
\begin{equation}
\begin{split}
\delta_2=& {}-\frac{2V}{U}V_{eff}'\left[2\frac{V'}{V}+\frac{V_{eff}''}{V_{eff}'}+\frac{\xi''}{\xi'}\right]\\
=&{}-\frac{2V}{U}V_{eff}'\left[\ln(V^2\xi'V_{eff}')\right]'.
\end{split}
\label{slrVeffd}
\end{equation}

So, $|\epsilon_1|\ll 1$ and $|\delta_1|\ll 1$ if $V_{eff}'$ is small enough. It allows us to use the effective potential for construction of the inflationary scenarios in models with the GB term.

Using the known formulae~\cite{Guo,Koh:2016abf} for  the tensor-to-scalar ratio $r$ and the spectral index $n_s$, we obtain:
\begin{equation}
\label{rVeff}
 r=8|2\epsilon_1-\delta_1|=\frac{4}{U}\left[\frac{d{\phi}}{dN}\right]^2=16\frac{V^2}{U^3}\left(V_{eff}'\right)^2,
\end{equation}
\begin{equation}
\label{nsVeff}
\begin{split}
   n_s=&1-2\epsilon_1-\frac{2\epsilon_1\epsilon_2-\delta_1\delta_2}{2\epsilon_1-\delta_1}=1-2\epsilon_1-\frac{d\ln(r)}{dN}\\
   =&1+\frac{d\ln(V/r)}{dN}=1+\frac{2}{U}\left(2V V_{eff}''+V' V_{eff}'\right).
\end{split}
\end{equation}

Moreover, the spectral index $n_s$ can be presented via derivatives of the effective potential only:
\begin{equation}
\label{nsVeefdN}
    n_s=1-\frac{d}{dN}\left(\ln\left(\frac{dV_{eff}}{dN}\right)\right).
\end{equation}

A standard way of the reconstruction of inflationary models~\cite{Mukhanov:2013tua,Koh:2016abf,Pozdeeva:2020shl} includes the assumption of explicit form of the inflationary parameter $n_s$ and $r$ as functions of $N$. Formula (\ref{nsVeefdN}) shows how the knowledge of $n_s(N)$ allows to calculate $V_{eff}(N)$.

The expression for amplitude $A_s$ in the leading order approximation is~\cite{vandeBruck:2015gjd}:
\begin{equation}
\label{As}
A_s\approx\frac{H^2}{\pi^2 U r}\approx\frac{V}{6\pi^2 U^2r}.
\end{equation}

In the slow-roll approximation, the e-folding number $N$ can be presented as the following function of $\phi$:
\begin{equation}
\label{N1}
N(\phi)=\int\limits^{\phi}_{\phi_{end}}\frac{dN}{d\phi}d\phi
\simeq\int\limits^{\phi_{end}}_{\phi}\frac{U}{2VV'_{eff}}d\phi.
\end{equation}
By this definition, $N<0$ during inflation. To get a suitable inflationary scenario we calculate inflationary parameters for $-65<N<-50$ and compare them with the observation data~\cite{Planck2018}.

Integrating Eq.~(\ref{EquphiloN}), one gets the function $\phi(N)$ is either in the analytic form, or in quadratures.
 We assume that $N=0$ at the end of inflation, and fix the value of the integration constant  by the condition $\epsilon_1(\phi(0))=1$.
After this, we know $\epsilon_i(N)$ and $\delta_i(N)$ and the inflationary parameters.

\subsection{Stability of de Sitter solutions}
The effective potential $V_{eff}$ is a useful tool to seek de Sitter solutions and to determine their stability~\cite{Pozdeeva:2019agu}.
In the case of the existence of a stable de Sitter solution, it is difficult to construct an inflationary scenario with a graceful exit. One might consider models with unstable de Sitter solutions or without an exact de Sitter solution, but  unstable quasi-de Sitter ones are more suitable.

In the case of a monomial potential $V$ and a more complicated function~$\xi$:
\begin{equation}
\label{monpot}
V=V_0\phi^n,\qquad \xi=\frac{3U^2}{V_0}\alpha\phi^q+\left(\beta+\frac{3U^2}{V_0}\right)\phi^{-n},
\end{equation}
with arbitrary constants $\alpha$, $\beta$, $V_0>0$, $n$ and $q\neq -n$, we have:
\begin{equation*}
 {V}_{eff}=\frac{U^2}{V_0\phi^n}\left(\alpha\phi^{q+n}+\frac{V_0\beta}{3U^2}\right),
\end{equation*}

In the case of $\alpha=0$ and $\beta\neq 0$, we obtain $\xi=C/V$, where $C=3U^2+V_0\beta$, and ${V}_{eff}$ is proportional to~$\xi$:
\begin{equation}
V_{eff}=\frac{C-3U^2}{3V}=\frac{\beta}{3\phi^{n}}
\end{equation}
and these is no de Sitter solution, because the effective potential ${V}_{eff}$ has no extremum for $n>0$. For $n<0$ there exist the only extremum for $\phi=0$, but the potential $V$ is not finite at  $\phi=0$. By the same reason, these is no de Sitter solution with $\phi_{dS}\neq 0$ in the case $\alpha\neq 0$ and $\beta= 0$.

In the case of nonzero $\alpha$ and $\beta$, the de Sitter point is given by
\begin{equation}
\label{phidS}
\phi_{dS}=\left(\frac{n\beta V_0}{3U^2\alpha q}\right)^{1/(q+n)}.
\end{equation}

In the present paper, we focus on the case of $q<0$ and $n>0$. The choice is motivated from the fact that,
in this case, the effective potential, which governs the slow-roll regime,
is flatter than the potential $V(\phi)$. We demonstrate later that it can improve
the situation with inflation in the simplest cases of massive and self-interacting
potentials. In the case of $q<0$ and $n>0$, we should assume that $\beta V_0/\alpha<0$ to get a de Sitter solution with a real $\phi_{dS}>0$.

At the de Sitter point, the second derivative of the effective potential is
\begin{equation}
{V}^{\prime\prime}_{eff}(\phi_{dS})=\frac{n\beta(q+n)}{3\phi_{dS}^{2+n}}.
\end{equation}
For positive values of $n$ and $\phi_{dS}$, the de Sitter solution is stable if $\beta(n+q)>0$ and unstable in the case $\beta(n+q)<0$.

The case of the quartic potential $V$ and a more complicated function $\xi$ will be investigated in Section~\ref{xi246}.

 In the next sections, we demonstrate the usefulness  of the effective potential for selected cases.

\section{Inflationary parameters in case of monomial potential with GB coupling}

\subsection{Application to the known model}
The choice of the function $\xi(\phi)=C/V(\phi)$, where $C$ is a constant, is actively studied~\cite{Guo:2009uk,Guo,Jiang:2013gza,Koh:2016abf,Yi:2018gse,Odintsov:2018,Kleidis:2019ywv,Rashidi:2020wwg,Pozdeeva:2020shl}. In this case,
\begin{equation}
    V_{eff}=\frac{C-3U^2}{3V},
\end{equation}
and the slow-roll parameters are as follows:
\begin{equation*}
    \epsilon_1=\frac{\left(3U^2-C\right){V'}^2}{3UV^2},\quad \epsilon_2=\frac{4\left(C-3U^2\right)\left(VV''-{V'}^2\right)}{3UV^2},
\end{equation*}
\begin{equation*}
    \delta_1=\frac{2C}{3U^2}\epsilon_1\,,\quad  \delta_2=\epsilon_2\,.
\end{equation*}
So, the inflationary parameters are
\begin{equation}
\label{InflationParamxiVm1ns}
    n_s=1+\frac{2\left(3U^2-C\right)\left(2VV''-3{V'}^2\right)}{3UV^2}\,.
\end{equation}
\begin{equation}
\label{InflationParamxiVm1r}
    r=\frac{16{V'}^2\left(3U^2-C\right)^2}{9U^3V^2},
\end{equation}

Note that in the case $C=3U^2$, the slow-roll approximation does not work, because all slow-roll parameters are identically equal to zero.

For $V=V_0\phi^n$, where $V_0$ and $n$ are constants, we integrate Eq.~(\ref{N1}), taking into account $\epsilon_1(\phi(0))=1$, and obtain
\begin{equation}
    \phi^2(N)=\frac{n(4N-n)(C-3U^2)}{3U}.
\end{equation}
So,
\begin{equation}
  \epsilon_1 ={}-{\frac {{n}}{4N-n}}, \quad \epsilon_2 ={}- {\frac {4}{4N-n}},
 \end{equation}
\begin{equation}
  \delta_1={}-{\frac{2C\,{n}}{3U^2(4N-n)}}, \quad \delta_2 ={}-\frac{4}{4N-n},
\end{equation}

From Eqs.~(\ref{InflationParamxiVm1ns}) and (\ref{InflationParamxiVm1r}), we obtain:
\begin{equation}
  n_s=1+\frac{2(n+2)}{4N-n}\,,  \quad r=\left|\frac{16n(C-3U^2)}{3U^2(4N-n)}\right|\,,
\end{equation}
\begin{equation}
A_s= \frac{V_0(4N-n)^{1+n/2}}{32\pi^2 (3U)^{n/2}[n(C-3U^2)]^{1-n/2}}\,.
\end{equation}

 In the case of a monomial potential $V$,  adding of the GB term with $\xi=C/V$ does not change $n_s$, but changes $A_s$ and $r$. Let us note, that for $n>2$, the combination $C-3U^2$ which is the numerator of the effective potential
enters in the numerators of both $A_s$ and $r$. This means that both the parameters can be made appropriately small if the corresponding effective potential is small, independently
of the magnitude of the actual potential $V(\phi)$, which allows us to obtain the required
values of spectral index and tensor-to-scalar ratio for the Higgs field, coupled appropriately
to the GB term. Note also, that in this scenario we need not introduce
a dimensionless parameter large in compared to unity similar to the theory of Higgs field
coupled to curvature~\cite{BezrukovShaposhnikov}. However, we yet need to address the problem associated with the spectral index $n_s$.

Substituting $-65<N<-50$, one can compare the inflationary parameters with the observation data.
In  case $n=4$ and $\epsilon_1= \epsilon_2= \delta_2$, we obtain in the slow-roll approximation
\begin{equation}
r=\left|\frac {16(C-3U^2)}{3U^2(N-1)}\right|,\qquad n_s=1+\frac{3}{N-1},
\end{equation}
The observation~\cite{Planck2018}: $n_s=0.9649\pm 0.0042$ at 68\% CL, implies that $-96<N<-75$. At the same time the number of e-foldings before the end of inflation at which observable perturbations were generated for the $\phi^4$ model without the GB term has been estimated~\cite{Liddle:2003as} as $64$ and we do not think that the addition of the GB term can essentially increase this number. By this reason, the model with the $\phi^4$ potential and $\xi\thicksim 1/\phi^4$ is ruled out.
In this paper, we find such a function $\xi(\phi)$ that the model with $V=V_0\phi^4$ does not contradict the observation data.
For $n>4$, we also get contradictions with the observation data.

We complete  the present subsection with a brief discussing on $n=2$ case, when both $n_s$, and $A_s$ do not depend on $C$. For $-65<N<-55$, one gets $0.9639<n_s<0.9695$ that is in a good agreement with observation.
Since  $\epsilon_2>\epsilon_1$ and $\delta_2>\epsilon_1$, then we cannot use the slow-roll approximation up to the point $N=0$, when $\epsilon_1=1$, but this approximation is valid for any $N<-1/2$ if $|C|< 3U^2$. Note that choosing $1.5U^2<C<4.5U^2$, one gets $r<0.0673$ that does not contradict the observation data. This means
that it is possible to revive a massive scalar field inflation, as stated in~\cite{Guo,Rashidi:2020wwg}.
Note, that unlike the $n>2$ case, the scalar field potential itself should ensure
the correct value of $A_s$, so the requirement that the mass of the scalar field should be of the
order of $10^{-5} M_{Pl}$ is  necessary.

\subsection{Models with monomial functions $V$ and $\xi$}

In the case of  $V=V_0\phi^n$ and $\xi=\xi_0\phi^{q}$, we get  the following expressions for the slow-roll parameters:
\begin{eqnarray}
\label{srp_eps1}
 \epsilon_1 &=&nU\frac{\alpha\,q{\phi}^{q+n}+n }{\phi^2}, \\
   \epsilon_2&=&2U\frac {2n-\alpha\,q{\phi}^{q+n} \left( n+q-2 \right)}{{\phi}^{2}},\\
  \delta_1&=&{}-2U\alpha\,q{\phi}^{q+n-2}\left(\alpha\,q{\phi}^{q+n}+n \right), \\
   \delta_2&=&{2U\frac {-2q\alpha\,{\phi}^{q+n} \left( n+q-1 \right) -n \left( n-2+q
 \right) }{{\phi}^{2}}}\label{srp_del2}
\end{eqnarray}
and for the inflationary parameters:
\begin{eqnarray*}
    n_s&=&1-\frac{2U}{\phi^2}\left([3n+2q-2]q\alpha\,\phi^{n+q}+n^2-2n\right),\\
     r&=&\frac{16U}{\phi^2}\left(q\alpha\,\phi^{n+q}+n\right)^2.
\end{eqnarray*}

The expression for amplitude is
\begin{equation}
\label{Asphi}
A_s\approx{\frac{{V_0}\,{\phi}^{n+2}}{12{\pi }^{2} \left(\alpha\,q {\phi}^{q+n}
+n \right) ^{2}}}.
\end{equation}

For $n=4$ and $q=-2$, the slow-roll parameters (\ref{srp_eps1})--(\ref{srp_del2}) can be presented in the following form:
\begin{eqnarray*}
 \epsilon_1&=&{}-\frac {8\,U \left( \alpha\,{\phi}^{2}-2 \right)}{\phi^2},\qquad
  \epsilon_2=\frac {16\,U}{{\phi}^{2}},\\
  \delta_1&=&{}-8\alpha\,U \left( \alpha\,\phi^2-2 \right),\qquad
   \delta_2=8\,\alpha\,U.
\end{eqnarray*}

The condition for the end of inflation  $\epsilon_1(\phi_{end})=1$ is satisfied
at the point $\phi_{end}=\sqrt{8/(1+4\,\alpha)}$.

For the considering case, the tensor-to-scalar ratio and the spectral index of scalar perturbations can be presented in the following form:
\begin{eqnarray}
n_s&=&1+8\,\alpha\,U-{\frac {48U}{{\phi}^{2}}},\label{ns4m2}\\
r&=&{\frac {64\,U \left( \alpha\,{\phi}^{2}-2 \right)^{2}}{{\phi}^{2}}}.
\label{r4m2}
\end{eqnarray}

The  expression for the amplitude is
\begin{equation}
A\approx{\frac {{V_0}\,{\phi}^{6}}{384{U}^{3}{\pi}^{2}
 \left( \alpha\,{\phi}^{2}-2 \right)^{2}}}\,.
\end{equation}

The e-folding number can be expressed  as follows:
\begin{equation}
N=\frac{\ln  \left( \left( 8\,\alpha\,U+1 \right)  \left(
2-\alpha\,{\phi}^{2} \right)/2 \right)}{8\alpha\,U},
\end{equation}
hence,
\begin{equation}
\phi^2=\frac {2\,(8\,\alpha\,U-{\mathrm{e}^{8U\alpha\,N}}+1)}{\alpha\,
 \left( 8\,\alpha\,U+1 \right) }
\,.
\end{equation}

The spectral index of scalar perturbations (\ref{ns4m2}) is given by the expression:
\begin{equation}
n_s={\frac { \left( 8\,\alpha\,U+1 \right)  \left(1-16\,\alpha\,U-{\mathrm{e}^
{8U\alpha\,N}}\right) }{8\,\alpha\,U-{\mathrm{e}^{8U\alpha\,N}}+1}},
\end{equation}
the tensor-to-scalar ratio (\ref{r4m2}) can be expressed as follows:
\begin{equation}
\begin{split}
r&={\frac {128\alpha\,U{\mathrm{e}^{16U\alpha\,N}}}{ \left( 8\,\alpha
\,U+1 \right)  \left(8\,\alpha\,U-{\mathrm{e}^{8U\alpha\,N}}+1\right) }}\\&= \frac{16(16U\alpha+n_s-1)^2}{3(8U\alpha-n_s+1)}\,.
\end{split}
\end{equation}

For $N={}-60$, the spectral index $n_s\approx0.9506$ at $\alpha=10^{-4}$.
For the same values of $\alpha$ and $N$ the amplitude is:
\begin{equation*}
A_s= \frac{V_0\left(8U\alpha-{\mathrm{e}^{8U\alpha\,N}}+1\right)^3}{192\pi^2U^3\alpha^3(8U\alpha+1)
{\mathrm{e}^{16U\alpha\,N}}}\approx62084.7\cdot V_0\,.
\end{equation*}

\begin{widetext}
\begin{figure}[h!tbp]
\includegraphics[width= 8cm, height= 8.3cm]{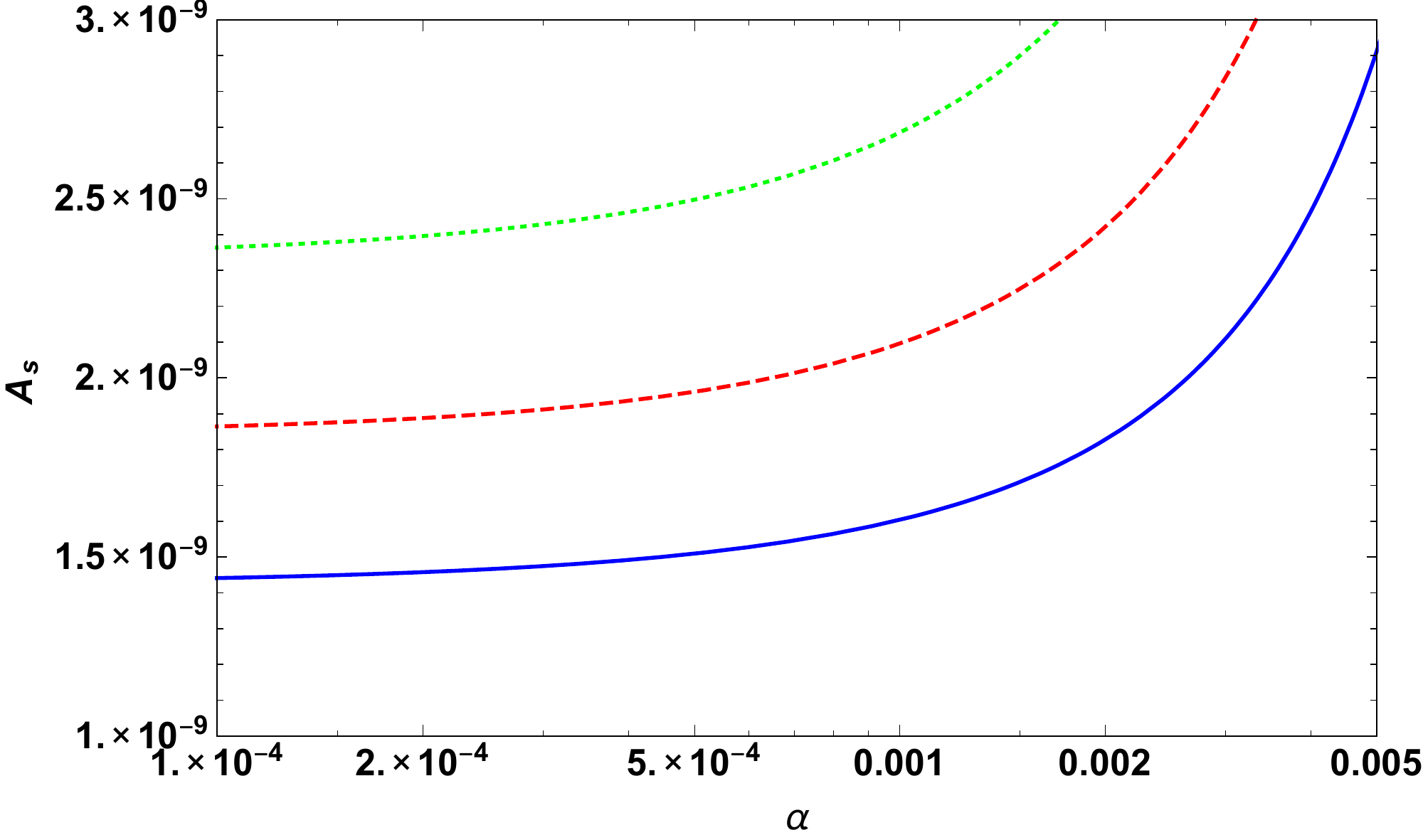}\,\,\,\,\,\,\includegraphics[width= 8cm, height= 8.3cm]{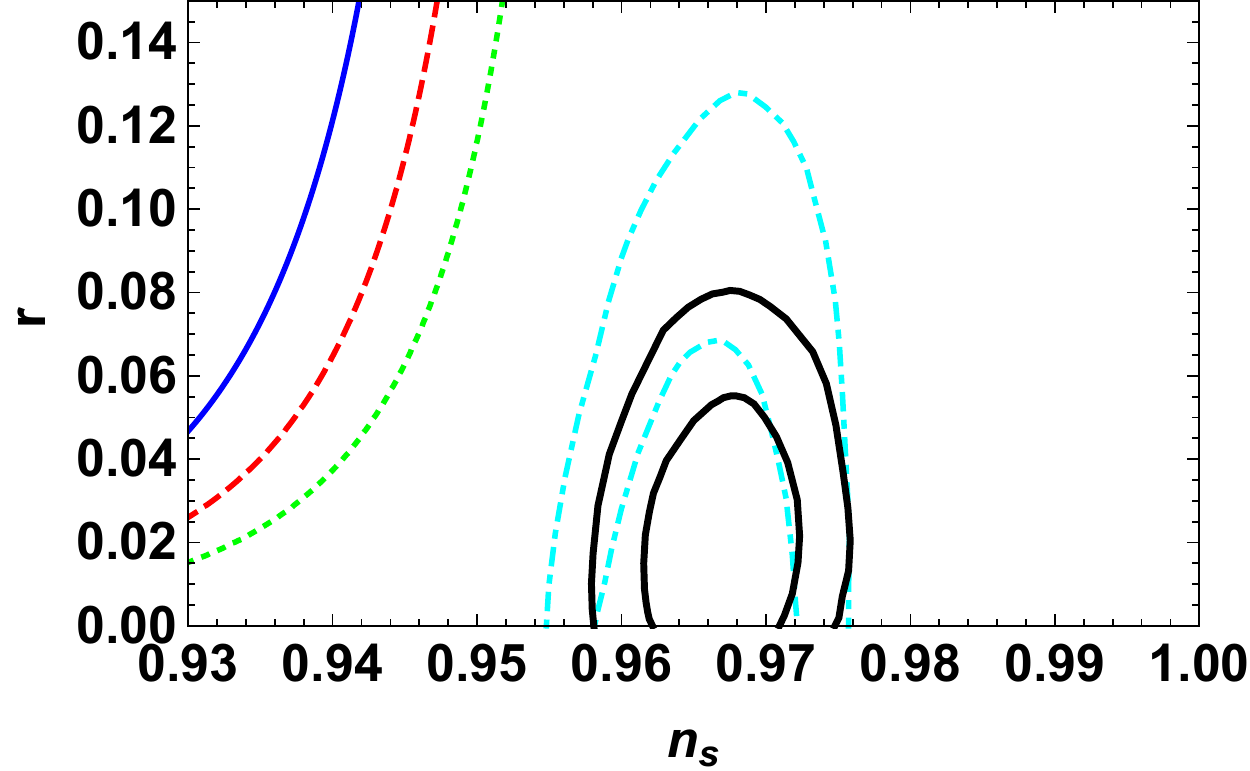}
\caption{Inflationary parameters for the model with $V=V_0\phi^4$ and $\xi=\xi_{2}\phi^{-2}$. Blue solid, red dashed and green dot curves correspond to $N=-55$, $N=-60$, and $N=-65$ respectively. We fix $U=1/2$. The contours correspond to the marginalized joint $68 \%$ and $95 \%$ CL. The cyan dash-dot lines are for the Planck'18 TT, TE, EE+ low E+ lensing data, whereas the black solid lines correspond to Planck'18 TT, TE, EE+ low E+ lensing + BK15+ BAO data. The pivot scale is fixed at usual $k= 0.002 MPc^{-1}.$
\label{f1}}
\end{figure}
\end{widetext}

 To get the required magnitude  of the amplitude $A_s$, we chouse the corresponding value of $V_0$. Substituting the same, we can find $n_s$ and $r$ as functions of $\alpha$.
The value of $n_s$ now does depend on the GB coupling, however, it
appears to be impossible to get both $n_s$ and $r$ in the observationally allowed
range (see Fig.~\ref{f1}).
This problem can be circumvented by considering
a more complicated form of the coupling function.
 This consideration
is the goal of the next section.

\section{The $\phi^4$ potential and complicated forms of the GB interaction}
\label{xi246}
In this section, we shall investigate models with quartic potential for the GB coupling function of a more complicated structure.

\subsection{The existence of de Sitter solutions}
Let us consider a more general model,
\begin{equation}
\label{three terms}
V=\lambda\phi^{4},\qquad \xi=\xi_2\phi^{-2}+\xi_4\phi^{-4}+\xi_6\phi^{-6},
\end{equation}
with arbitrary constants $\xi_2$, $\xi_4$, and $\xi_6$. The dimensionless constant $\lambda$ in the potential will be fixed to $\lambda=0.1$  in order to describe a large field approximation for the Higgs potential.

The condition $V_{eff}^\prime(\phi_{dS})=0$ gives the following values of $\phi_{dS}^2$:
\begin{equation}\label{phidSpos}
    \phi^2_{dS} =
    \left\{
    \begin{split}
    \frac{-\beta\pm\sqrt{\beta^2-3\xi_2\xi_6}}{\xi_2},&\quad\mbox{at}\quad \xi_2\neq 0,\\
    {}-\frac{3\xi_6}{2\beta},&\quad\mbox{at}\quad \xi_2= 0,
    \end{split}
    \right.
\end{equation}
where $\beta=\xi_4-3U^2/\lambda$.

 Without loss of generality, we consider $\phi_{dS}>0$ only\footnote{The case $\phi_{dS}=0$ is excluded because the function $\xi$ is singular at $\phi=0$.}.
The condition $\phi^2_{dS}>0$ gives restrictions on values of the parameters.
If $\xi_2= 0$, then a de Sitter solution exists if and only if $\xi_6/\beta<0$.
In the case $\xi_2\neq 0$, we have the following possibilities:
\begin{enumerate}
\item If $\xi_6\xi_2>\beta^2/3$, then these is no de Sitter solution.
\item If $0\leqslant\xi_6\xi_2\leqslant\beta^2/3$ and $\xi_2\beta>0$, then these is no de Sitter solution.
\item If $0<\xi_6\xi_2<\beta^2/3$ and $\xi_2\beta<0$, then there exist two de Sitter solutions.
\item If $\xi_6\xi_2=\beta^2/3$ and $\xi_2\beta<0$, then there exists only one de Sitter solution.
\item If $\xi_6\xi_2<0$, then there exists only one de Sitter solution.
\item If $\xi_6=0$ and $\xi_2\beta<0$, then there exists only one de Sitter solution.
\end{enumerate}

The stability of a de Sitter point is defined by the sign of $V_{eff}''(\phi_{dS})$. In particular,
if $\xi_2= 0$, then
\begin{equation}\label{D2VeffdS}
    V_{eff}''(\phi_{dS})=\frac{64\beta^4}{81\xi_6^3},
\end{equation}
so a de Sitter solution is stable at $\beta<0$ and $\xi_6>0$  and is unstable at $\beta>0$ and $\xi_6<0$.
If $\xi_6= 0$, then de Sitter solution is stable at $\beta>0$ and $\xi_2<0$ and is unstable at $\beta<0$ and  $\xi_2>0$.
To get a suitable inflationary scenario we can consider both the case with unstable de Sitter solution, and the case without de Sitter solution.

\begin{widetext}
\subsection{The inflationary parameters}
For the  model under consideration, we obtain the following expressions,
\begin{equation}\label{epsV4xi246}
    \epsilon_1={}-\frac{8\lambda}{3U\phi^4}\left(\xi_2\phi^4+2\beta\phi^2+3\xi_6\right),\quad  \epsilon_2={}-\frac{16\lambda}{3U\phi^4}\left(\beta\phi^2+3\xi_6\right),
\end{equation}
\begin{equation}\label{delta1V4xi246}
    \delta_1={}-\frac{8\lambda}{9U^3\phi^6}\left(\lambda\phi^4\xi_2+(6U^2+2\lambda\beta)\phi^2+3\lambda\xi_6\right)\left(\phi^4\xi_2+2\beta\phi^2+3\xi_6\right),
\end{equation}
\begin{equation}\label{delta2V4xi246}
    \delta_2={}-\frac{8\lambda(-\lambda\xi_2^2\phi^8+2(6U^2\beta+2\lambda\beta^2+3\lambda\xi_2\xi_6)\phi^4+12\xi_6(3U^2+2\lambda\beta)\phi^2+27\lambda\xi_6^2)}
{3U\phi^4(\lambda\xi_2\phi^4+6U^2\phi^2+2\lambda\beta\phi^2+3\lambda\xi_6)}.
\end{equation}

The inflationary parameters are as follows:
\begin{equation}
\label{nsrV4xi246}
    n_s=1+\frac{8\lambda(\xi_2\phi^4+6\beta\phi^2+15\xi_6)}{3U\phi^4},\qquad r=\frac{64\lambda^2\left(\xi_2\phi^4+2\beta\phi^2+3\xi_6\right)^2}{9U^3\phi^6}.
 \end{equation}
\end{widetext}

\begin{figure}[h!tbp]
\centering
\includegraphics[width= 8cm, height= 7cm]{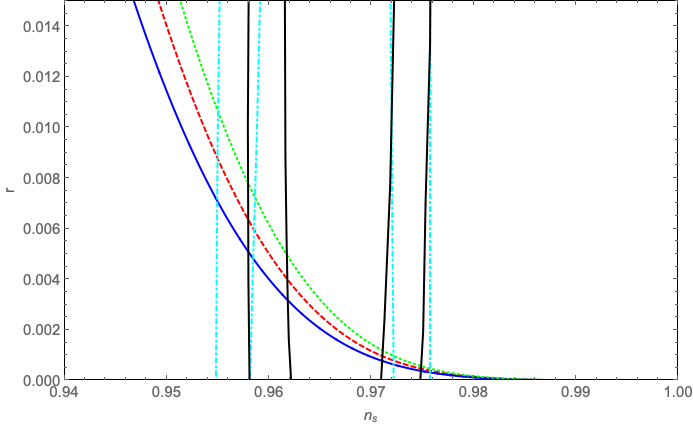}
\caption{Parameter space of $n_s$ and $r$ for the model with $V=\lambda\phi^4$ and $\xi=\xi_{2}\phi^{-2}+\xi_{4}\phi^{-4}+\xi_{6}\phi^{-6}$ in the case
$\beta \neq0$. Blue solid, red dashed and green dot curves correspond to $N=-55$, $N=-60$, and $N=-65$ respectively. The contours correspond  to the marginalized joint $68 \%$ and $95 \%$ CL. The cyan dash-dot lines are for the Planck'18 TT, TE, EE+ low E+ lensing data, whereas  the black solid lines correspond  to Planck'18 TT, TE, EE+ low E+ lensing + BK15+ BAO data. The pivot scale is fixed at usual $k= 0.002 MPc^{-1}.$
\label{f2}}
\end{figure}

  In the generic case, when  parameters $\xi_2$, $\xi_6$, and $\beta$ are nonzero, analytical solutions cannot be obtained. Thus numerical methods are being implemented to get the relationships between the inflationary parameters. In Fig.~\ref{f2}, one can see that for the e-foldings number $N$ between $-65$ and $-55$, one can always get $n_s$ and $r$ in the observational range for particular choice of $\xi_2$. Here, we have taken $\xi_6=-0.1$ and $U=1/2$. The value of $\beta$ is taken to be $\beta= -7.4$.  The choice of $\beta$ is from the fact to keep $A_s \sim 2.1\times 10^{-9}$. The parameter $\xi_2$ is taken in the range $0\leq\xi_2\leq 0.5$.

It is also interesting to consider a few particular cases, when parameters $\xi_2$, $\xi_6$, or $\beta$ are equal to zero. In these cases some analytic results can be obtained. In particular, if $\xi_6=0$ and $\beta=0$, then $\epsilon_1$ is a constant, hence, the slow-roll inflation is not possible in this case.  We consider other particular cases in the next subsections of this section.

\subsection{The case of $\beta=0$}

In the case of $\xi_6\neq0$ and $\beta=0$,
the inflationary parameters (\ref{nsrV4xi246}) are as follows:
\begin{equation}
\label{nsrbeta0}
\begin{split}
    n_s&=1+\frac{8\lambda(\xi_2\phi^4+15\xi_6)}{3U\phi^4},\\
     r&=\frac{64\lambda^2\left(\xi_2\phi^4+3\xi_6\right)^2}{9U^3\phi^6}.
 \end{split}
 \end{equation}

The solution of Eq.~(\ref{EquphiloN}) with an additional condition $\epsilon_1(\phi(0))=1$ has the following form:
\begin{equation}\label{phiNbeta0}
    \phi(N)=\sqrt[4]{\frac{3\xi_6(3U\mathrm{e}^{16\lambda\xi_2N/(3U)}-8\lambda\xi_2-3U)}{\xi_2(8\lambda\xi_2+3U)}}
    \end{equation}

\begin{figure}[h!tbp]
\centering
\includegraphics[width= 8cm, height= 7cm]{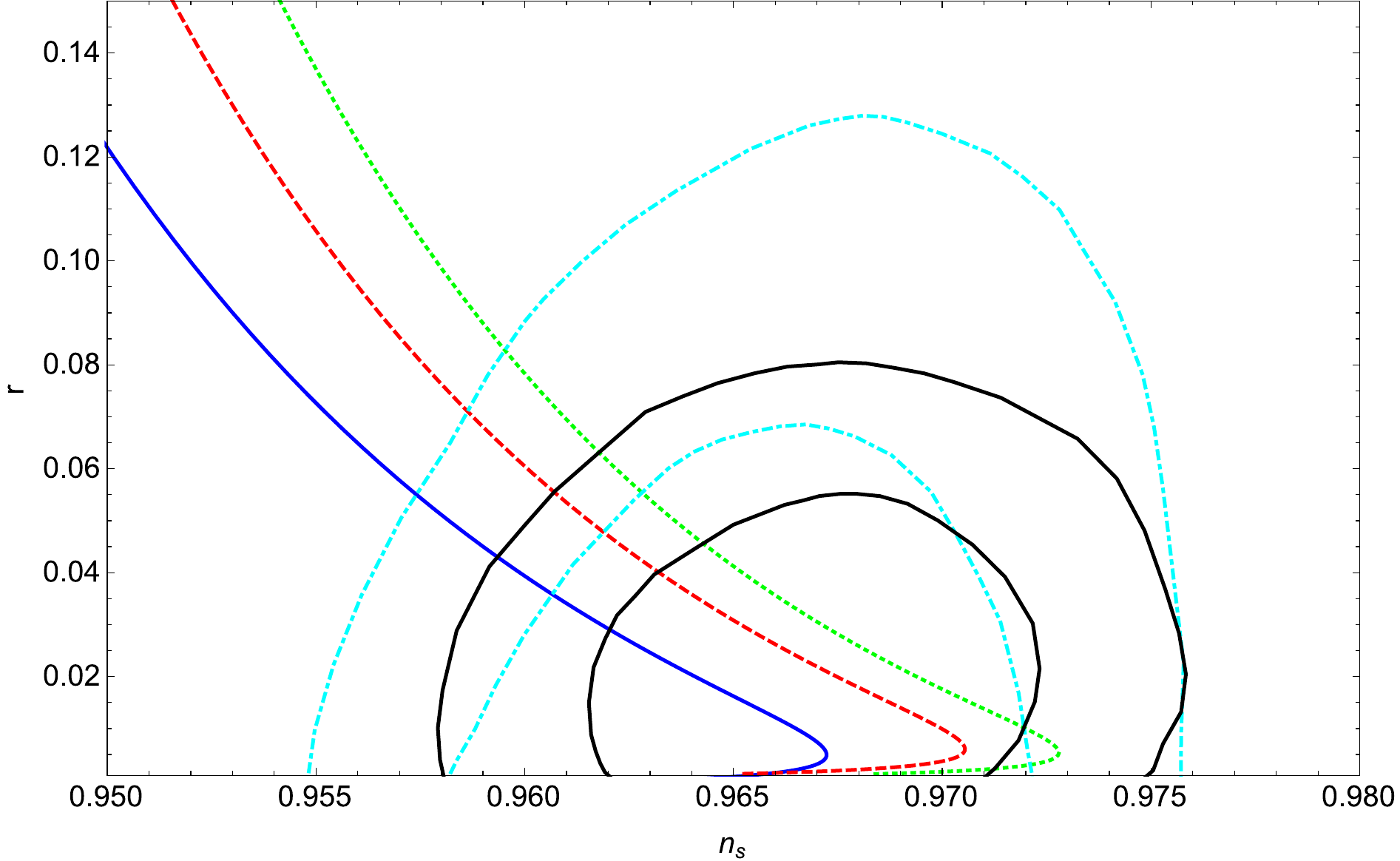}
\caption{Parameter space of $n_s$  and $r$ for the model with $V=\lambda\phi^4$ and $\xi=\xi_{2}\phi^{-2}+\xi_{4}\phi^{-4}+\xi_{6}\phi^{-6}$ in the case
$\beta=0$. Blue solid, red dashed, and green dot curves correspond to $N=-55$, $N=-60$, and $N=-65$ respectively. The contours  correspond  to the marginalized joint $68 \%$ and $95 \%$ CL. The cyan dash-dot lines are for the Planck'18 TT, TE, EE+ low E+ lensing data, whereas  the black solid lines correspond to Planck'18 TT, TE, EE+ low E+ lensing + BK15+ BAO data. The pivot scale is fixed at usual $k= 0.002 MPc^{-1}.$
\label{f3}}
\end{figure}

\begin{widetext}
Substituting~(\ref{phiNbeta0}) into expressions~(\ref{nsrbeta0}), we get:
\begin{equation}
    n_s= 1+\frac{8\lambda\xi_2(3U\mathrm{e}^{16\lambda\xi_2N/(3U)}+32\lambda\xi_2+12U)}{3U(3U\mathrm{e}^{16\lambda\xi_2N/(3U)}-8\lambda\xi_2-3U)},
\end{equation}
\begin{equation}
r=\frac{64\sqrt{3}\lambda^2\xi_2^3\xi_6\mathrm{e}^{32\lambda\xi_2N/(3U)}\sqrt{(8\lambda\xi_2+3U)}}
{U(3U\mathrm{e}^{16\lambda\xi_2N/(3U)}-8\lambda\xi_2-3U)(8\lambda\xi_2+3U)\sqrt{\xi_2^3\xi_6(3U\mathrm{e}^{16\lambda\xi_2N/(3U)}-8\lambda\xi_2-3U)}}
\end{equation}

The inflationary parameter $n_s$ does not depend on $\xi_6$, but we cannot put $\xi_6=0$, because $\phi(N)\equiv 0$ in this case. In Fig.~\ref{f3}, we show the variation of $r$ with $n_s$ for a particular choice of $\xi_6= -0.1$. For lower value of $\xi_6$ the value of $r$ goes lower.
\end{widetext}

\subsection{The case $\xi_6=0$ and $\beta\neq 0$}

In the case $\xi_6=0$ the inflationary parameters (\ref{nsrV4xi246}) are
\begin{equation}\label{Infparamxi2xi4}
n_s = 1+\frac{8\lambda}{3U}\xi_2+\frac{16\lambda\beta}{U\phi^2}, \qquad
  r =\frac{64\lambda^2(\xi_2\phi^2+2\beta)^2}{9U^3\phi^2}\,.
\end{equation}

We fix $\phi_{end}\equiv\phi(0)={}-(16\beta \lambda)/(8\xi_2\lambda+3U)$  by the condition $\epsilon_1=1$, solve Eq.~\eqref{N1} and get:
\begin{equation}
\label{phi_d}
 \phi^2=B\left(\frac{\mathrm{e}^{A_1N}}{A_1+1}-1\right),
  \end{equation}
 \begin{eqnarray*}
  n_s&=&1+A_1+\frac{3BA_1}{\phi^2}=1+A_1+\frac{3A_1(A_1+1)}{\mathrm{e}^{A_1N}-A_1-1} \\
  r&=&\frac{A_1^2(\phi^2+B)^2}{U\phi^2}=\frac{A_1^2B\mathrm{e}^{2A_1N}}{U({A_1+1})
  \left(\mathrm{e}^{A_1N}-A_1-1\right)} \\
  A_s&=&\frac{V}{6\pi^2U^2r}=\frac{\lambda B}{6U\pi^2A_1^2}\frac{\left({\mathrm{e}^{A_1N}}-A_1-1\right)^3}{(A_1+1)
  {\mathrm{e}^{2A_1N}}}
  \end{eqnarray*}
where $A_1=8\lambda\xi_2/(3U)$ and $B=2\beta/\xi_2$.

To get appropriate  values for inflationary parameters we suppose $N=-65$ is a start point of inflation,
 $\lambda=0.1$, $U=1/2$, $A_1=-0.01517$, and $B=2\cdot10^{-10}$:
\begin{equation}
n_s\approx0.9584,\quad r\approx 3.96\cdot10^{-13},\quad A_s\approx 2.02\cdot10^{-9}.
\end{equation}

\begin{widetext}
\begin{figure}[h!tbp]
\includegraphics[width= 8cm, height= 8cm]{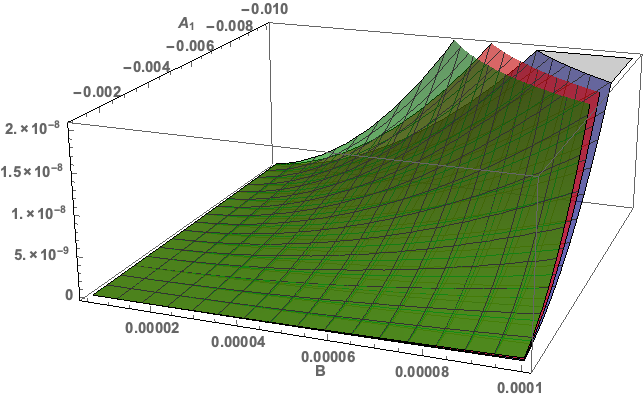}\,\,\,\,\,\includegraphics[width= 8cm, height= 8cm]{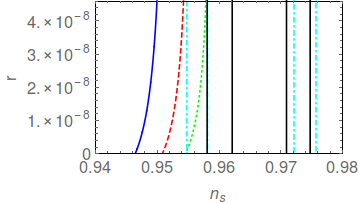}
\caption{Inflationary parameters for the model with $V=\lambda\phi^4$ and $\xi=\xi_{2}\phi^{-2}+\xi_{4}\phi^{-4}$ in the case
$\beta\neq 0$. In the left picture, blue plane corresponds to $N=-55$, red plane corresponds to  $N=-60$,
and green plane correspond to $N=-65$ [from the bottom upwards].
 The z-axis corresponds to $A_s$. In the right picture, the contours  correspond  to the marginalized joint $68 \%$ and $95 \%$ CL.
 Blue solid, red dashed, and green dot curves correspond to $N=-55$, $N=-60$, and $N=-65$ respectively.
 The cyan dash-dot lines are  for the Planck'18 TT, TE, EE+ low E+ lensing data, whereas the black solid lines correspond  to Planck'18 TT, TE, EE+ low E+ lensing + BK15+ BAO data.
 The pivot scale is fixed at usual $k= 0.002 MPc^{-1}$.
\label{f4}}
\end{figure}
\end{widetext}

Thus, if $\xi_2=-0.0284$ and $\beta=-2.84\cdot10^{-12}$, then we get appropriate inflationary parameters at $N=-65$.

\subsection{The case $\xi_2=0$}

The behaviour of the inflationary parameters for this case is given in the Fig.~\ref{f4}. Thus, from Fig.~\ref{f4}, it is obvious the observationally best suited case corresponds to $N=-65$ presented by the green line (plane).
\begin{widetext}
\begin{figure}[h!tbp]
\includegraphics[width= 8cm, height= 8cm]{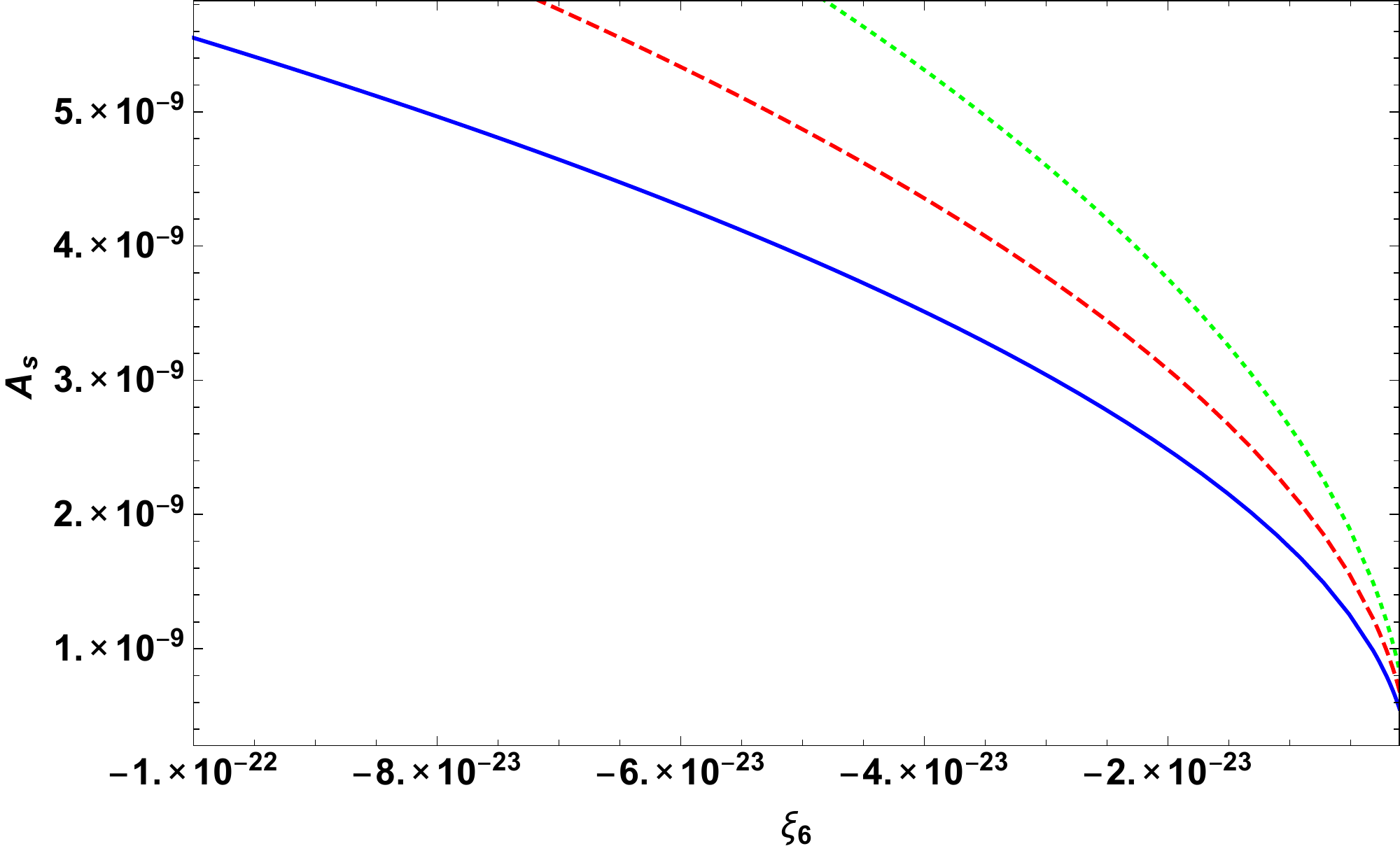}\,\,\,\,\,\includegraphics[width= 8cm, height= 8cm]{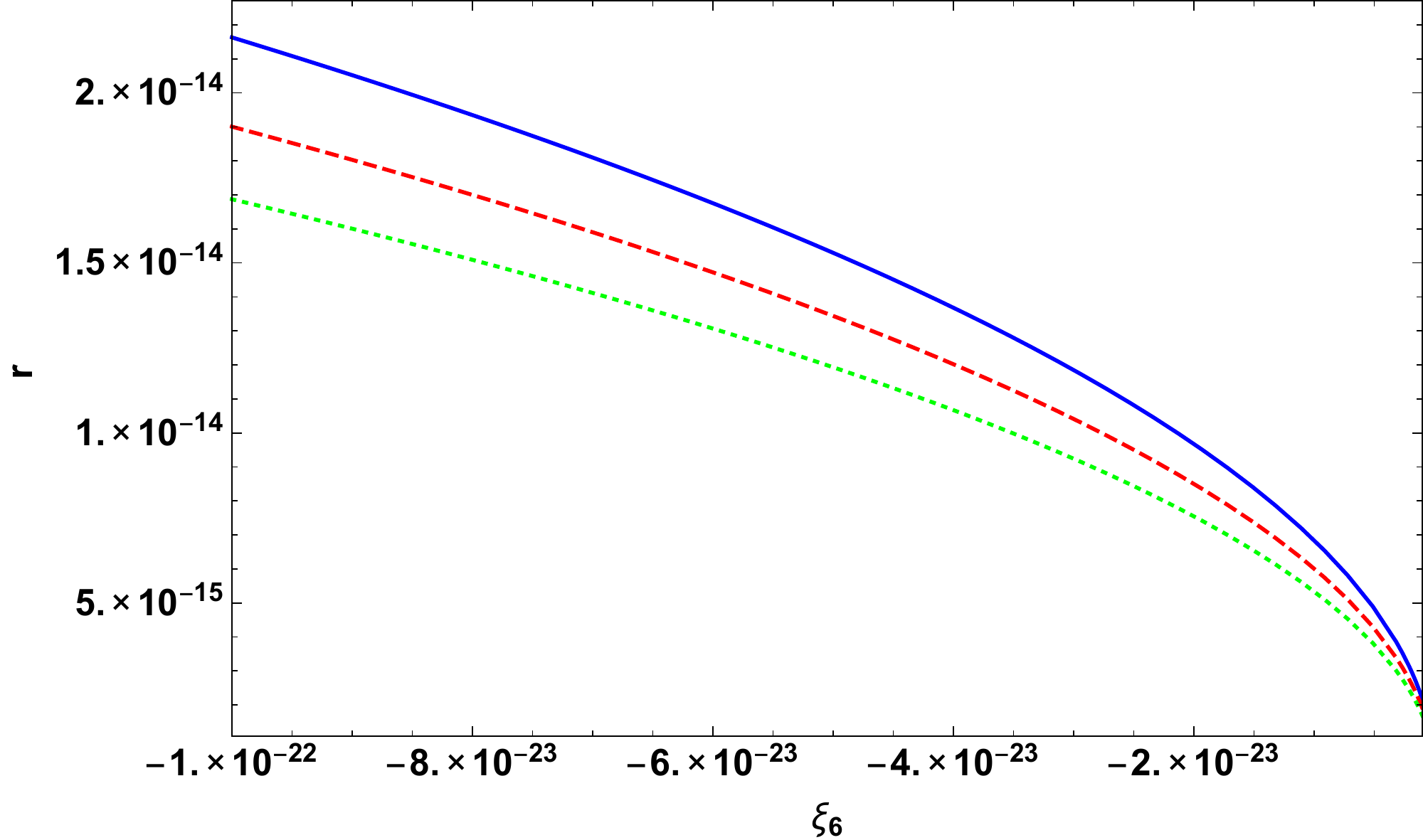}
\caption{Inflationary parameters $A_s$ (left) and $r$ (right) for the model with $V=V_0\phi^4$ and $\xi=\xi_{4}\phi^{-4}+ \xi_{4}\phi^{-6}$. Blue solid, red dashed and green dot curves correspond to $N=-55$, $N=-60$, and $N=-65$ respectively.
\label{fig2}}
\end{figure}
\end{widetext}

In the case of $\xi_6>0$, there  exists a stable de Sitter solution that can make such models unsuitable for the description of the early Universe evolution.
In the case of $\xi_6<0$, one gets the following inflationary parameters:
\begin{eqnarray}
  r&=&{\frac {64{{\lambda}}^{2} \left( 2\,\beta\,{\phi}^{2}+3
\,{\xi_6}\right)^{2}}{9{\phi}^{6}{U}^{3}}}, \\
  n_s&=&1+{\frac {16\,\beta\,{\lambda}}{U{\phi}^{2}}}+{\frac {40\,{\lambda}\,{
\xi_6}}{U{\phi}^{4}}},\\
A_s&=&{\frac {3U{\phi}^{10}}{128{\lambda}\,{\pi}^{2} \left( 2\,
\beta\,{\phi}^{2}+3\,{\xi_6} \right)^{2}}}.
\end{eqnarray}

To get the following relation between the e-folding number $N$ and the field $\phi$:
\begin{eqnarray}
N&\simeq&{}-\int\limits^{\phi_{end}}_{\phi}\frac{3U\phi^3}{4\lambda(2\beta\phi^2+3\xi_6)}d\phi
\nonumber\\
\label{N(phi)beforesl}
&=&\frac{3U{\phi}^{2}}{16\beta\,{\lambda}}-\frac
{9U{\xi_6}\,\ln\left( 2\,\beta\,{\phi}^{2}+3\,{\xi_6} \right) }{32{\beta}^{2}{\lambda}}+\frac {4\,\beta\,{\lambda}-\sqrt {-2{\lambda}\left( -8\,{\lambda}{\beta}^{2}+9\,U{\xi_6}\right) }}{8\beta{\lambda}}\\
&+&\frac{9U\xi_6}{32{\beta}^{2}{{\lambda}}}\ln\left( 3\,{\xi_6}+\frac { 4\left( -4\,\beta\,{\lambda}+\sqrt {-2\,{\lambda}\,
 \left( -8{\lambda}{\beta}^{2}+9\,U{\xi_6} \right) } \right)
\beta}{3U} \right) ,\nonumber
\end{eqnarray}
we use Eq.~(\ref{N1}) and define the end point of inflation using the condition $\epsilon_1(\phi=\phi_{end})=1$:
\begin{equation*}
\phi_{end}^2=\frac{2}{3U}\left(-4\,\beta\,\lambda+\sqrt{16\lambda^2\beta^2-
18U\lambda\xi_6}\right).
\end{equation*}
Note that this result is correct for $\beta\neq 0$ only. The case $\beta=0$ is considered separately.

The inverse [in context of \eqref{N(phi)beforesl}]  relation between field and e-folding number is rather complicated and requires numerical  considerations. However in particular case $\xi_4={\frac {3{U}^{2}}{{\lambda}}}$ ($\beta=0$), all
expressions can be simplified including the relations between field and e-folding number and vice versa:
\begin{equation}
  N=\frac {{\phi}^{4}U}{16{\lambda}\,{\xi_6}}+\frac12,\quad \phi^4={\frac{ 8{\lambda}\,{\xi_6}}{U}}\left(2N-1 \right).
\end{equation}

Therefore, in the case $\beta=0$, we get the following inflationary parameters
\begin{equation}\label{nsr6}
    n_s= 1+\frac{5}{2N-1},\quad r=\frac{2\sqrt{2\lambda}\xi_6}{\sqrt{U^3\xi_6(2N-1)^3}},
    \end{equation}
    and
    \begin{equation}
     A_s=\frac{\sqrt {2} \left( {{ \left( 2\,N-1 \right) {\lambda}\,{\xi_6}}} \right)^{5/2}}{3{\pi }^{2}U^{3/2}{\lambda}{{\xi_6}}^{2}}.
\end{equation}

In this case, the scalar spectral index is independent of $\xi_6$. In the range, $-65\leq N\leq -55$, the range of $n_s$ is $0.961832\geq n_s\geq 0.954955$. As usual in this case, we put $U= 1/2$ and $\lambda= 0.1$. The variation of $A_s$ and $r$ with respect to $\xi_6$ is plotted in Fig.~\ref{fig2}.

\section{Conclusions}

In this paper, we have studied inflation in models with the GB coupling. The presence of the GB coupling crucially modifies the underlying dynamics of the system~\cite{Tsujikawa:2006ph}.
As demonstrated in the earlier work~\cite{Pozdeeva:2019agu}, the existence and stability of de Sitter solution in such models can be analysed in terms of
the effective potential solely. As for the inflationary parameters, it appears that the effective potential is
not enough to characterize them fully. Nevertheless, such a notation can be helpful to understand
how these parameters vary in different models since the effective potential enters in the most
formulas derived to determine them. We have considered several examples of the power-law scalar field potential $V(\phi)$ and the coupling function $\xi(\phi)$. In the case of $V = V_0 \phi^n$ and $\xi = \xi_0 \phi^{-n}$, the effective potential can be made arbitrary small by choosing appropriate coupling coefficients. We show that under this condition, both the amplitude of scalar perturbations and the tensor-to-scalar ratio can be turned to arbitrary small numbers if $n>2$. Note, that this does not require smallness of $V_0$.  This means that the amplitude can be turned small required by observations even for the Higgs field, since the Higgs potential can be approximated as a monomial one with $n=4$ and $V_0=0.1$ for large $\phi$. Note that in this case, we have
not much freedom in choosing the coupling function --- since the effective potential is a difference of two functions, one of which is proportional to the coupling function, and the second is inversely proportional
to the potential, we should choose the coupling function to be almost equal to the inverse potential.
An exact cancellation of these two terms in the effective potential does not give us a viable model since the scalar field dynamics is absent in this case. In the case of ''almost cancellation'' the amplitude of scalar
perturbations can be made small for the Higgs field.
However, the spectral index $n_s$, which
does not depend upon $\xi_0$ in this case, turns out to narrowly missing the range of values admitted by observation for $N<66$ e-folds.

If we abandon the assumption that the coupling function has the power index inverse to that of the potential, the effective potential cannot be made small in general for large $V$ and $\xi$. This means that we cannot use such models for describing inflation with the potential too large to ensure the required small values of $A_s$ similar to the case of the Higgs-driven inflation. However, if appropriate $A_s$ is already ensured by the potential itself
(for $V=\lambda \phi^4$ this requires a very small $\lambda$), the coupling to GB term can improve
the situation with the tensor-to-scalar ratio $r$. In this case, we have more freedom in the coupling function,
though our choice is still restricted since we do not want for the slow-roll regime to end in a stable de Sitter solution. For a power-law potential proportional to $\phi^n$ and a positive coupling function proportional
to $\phi^{-m}$ we need $n>m$. An unstable de Sitter solution exists for such choice, and as
$V_{eff}'=0$ in any de Sitter point,
 the expression (\ref{rVeff}) indicates that the tensor-to-scalar ratio can be made sufficiently small. We have considered a particular case of quartic potential ($n=4$) with the $m=2$ coupling function.
   The classic inflationary scenario for a minimally coupled scalar field with a quartic potential is now ruled out due to inappropriately large value of $r$, the problem can be addressed by invoking the GB coupling. The main remaining hurdle of the quartic potential with nonminimal coupling to the GB term is related to $n_s$ which can be made consistent with observation only for coupling giving rise to inappropriately large~$r$ (see Fig.~\ref{f1}).

We show that the aforesaid difficulties can be circumvented by considering a more complicated form of the coupling function.
We use
\begin{equation}
\label{xigen}
 \xi=\xi_2 \phi^{-2} + \xi_4 \phi^{-4} + \xi_6 \phi^{-6}
\end{equation}
and demonstrate that the respective model fits the observational data~\cite{Planck2018} for all three considered observables: $A_s$, $n_s$, and $r$. An appropriate choice of the coefficients allows us to obtain this fit for the Higgs field (see Figs.~\ref{f2} $\&$ \ref{f3}). The chosen form of the function $\xi(\phi)$ has a singularity at $\phi=0$ that may arise difficulties for the consideration the evolution of the Universe after inflation. This problem can be solved by the introduction of a small positive constant $M_*$ and the consideration of the function
\begin{equation}
\label{xigenmod}
\xi_*=\frac{\xi_2}{M_*^2+\phi^{2}} + \frac{\xi_4}{\left(M_*^2+\phi^{2}\right)^{2}} + \frac{\xi_6}{\left(M_*^2+\phi^{2}\right)^{3}},
\end{equation}
where the value of the constant $M_*\ll\phi_{end}$. So, during inflation the function (\ref{xigen}) is a good approximation for the function (\ref{xigenmod}), whereas after inflation, when $\phi\ll M_*$, this function is almost a constant and the behaviour of this cosmological model is similar to the corresponding model without the GB term.

Another possible way out to improve upon $\phi^4$ model could be provided by  invoking a usual nonminimal coupling with the curvature $R$. The corresponding models without the GB term are studied in Refs.~\cite{Barvinsky:1994hx,Cervantes-Cota1995,BezrukovShaposhnikov}; the framework is extended to the Einstein-Gauss-Bonnet gravity in Ref.~\cite{vandeBruck:2015gjd}. It would be interesting to investigate  the model with generic forms of coupling functions and we defer  the same to our future study.

{ \ }

\section*{Acknowledgements}

This work is partially supported by Indo-Russia Project: E.P., A.T., and S.V.~are supported by the RFBR grant 18-52-45016 and M.S.~is supported by INT/RUS/ RFBR/P-315. A.T.~is supported by Russian Government Program of Competitive Growth of Kazan Federal University. Work of M.R.G.~is supported by the Department of Science and Technology, Government of India under the Grant Agreement number IF18-PH-228 (INSPIRE Faculty Award). M.R.G.~wants to thank Nilanjana Kumar for useful discussions regarding numerical solutions.

\end{document}